\documentclass{PoS}

\title{The repulsive core of the NN potential and the operator product expansion}

\ShortTitle{The repulsive core of the NN potential and the operator product expansion}

\author{\speaker{Sinya Aoki}\\
        Graduate School of Pure and Applied Sciences, University of Tsukuba, Tsukuba, Ibaraki 305-8571, Japan\\
        E-mail: \email{saoki@het.ph.tsukuba.ac.jp}}

\author{Janos Balog\\
        Research Institute for Particle and Nuclear Physics, 1523 Budapest 114, Pf. 49, Hungary\\
        E-mail: \email{balog@rmki.kfki.hu}}
 \author{Peter Weisz\\
        Max-Planck-Institut f\"ur Physik, F\"oringer Ring 6, D-80805 M\"unchen, Germany\\
        E-mail: \email{pew@mppmu.mpg.de}}       

\abstract{We investigate the short distance behavior of the 
nucleon--nucleon (NN) potential defined through the Bethe-Salpeter wave 
function, by perturbatively calculating anomalous dimensions of 6--quark 
operators in QCD. Thanks to the asymptotic freedom of QCD, 
the 1-loop estimations give exact results 
for the potential in the zero distance limit. We show that the 
chiral symmetry of the gauge interaction implies the existence of an operator 
whose anomalous dimension is zero for a given quantum number.
Furthermore we find that non-zero anomalous dimensions of other operators 
are all negative. These results predict the functional form of the NN 
potential at short distance, which is a little weaker than $r^{-2}$.  
On the other hand, the computation of the anomalous dimension spectrum alone 
can not determine whether the potential is repulsive or attractive at 
short distance. An additional analytic non-perturbative analysis suggests 
that the force at short distance is indeed repulsive at low energy as found
numerically. Some extensions of the method are briefly discussed.}

\FullConference{The XXVII International Symposium on Lattice Field Theory - LAT2009\\
		 July 26-31 2009\\
		 Peking University, Beijing, China}

\begin{document}

\section{Introduction}
In a recent paper \cite{IAH}, the nucleon-nucleon(NN) potentials have been 
calculated in lattice QCD from the Bethe-Salpeter (BS) wave function 
through the Schr\"odinger equation.
The results qualitatively resemble phenomenological NN potentials which 
are widely used in nuclear physics. The force at medium to long distance 
($r\ge 2$ fm) is shown to be attractive. This feature has long well been 
understood in terms of pion and other heavier meson exchanges.
At short distance, a characteristic repulsive core is reproduced by the 
lattice QCD simulation \cite{IAH}. No simple theoretical explanation, 
however, exists so far for the origin of the repulsive core. 

A hint to understand the repulsive core theoretically has appeared in 
Ref.\cite{ABW}, where properties of the BS wave function of the Ising 
field theory in 2-dimensions,
\begin{eqnarray}
\varphi(r,\theta) = i \langle 0 \vert \sigma(x,0)\sigma(0,0)\vert \theta, -\theta\rangle^{\rm in},
\quad r=\vert x\vert
\end{eqnarray}
are considered analytically. Here $\theta$ is the rapidity of the one 
particle state. From the operator product expansion (OPE) 
\begin{eqnarray}
\sigma(x,0)\sigma(0,0)\mathop{\sim}_{r\to0} G(r){\bf 1} 
+ c\, r^{3/4}\, {\cal E} (0) + \cdots,
\end{eqnarray}
the BS wave function at short distance becomes
\begin{eqnarray}
\varphi(r,\theta) &\sim& C\, r^{3/4} \sinh(\theta) + O(r^{7/4}),
\end{eqnarray}
which predicts the short distance behavior of the potential as
\begin{eqnarray}
V_\theta (r) &=& \frac{ \varphi^{''}(r,\theta) + \sinh^2\theta \varphi(r,\theta)}{\varphi(r,\theta)}
\sim -\frac{3}{16}\frac{1}{r^2}.
\end{eqnarray}
The OPE in this case predicts not only the $r^{-2}$ behavior of the 
potential at short distance but also its coefficient $-3/16$. 
The potential at short distance does not depend on the energy 
(rapidity) of the state; it is universal.

In this report, the OPE analysis is applied to QCD, with the aim 
to theoretically better understand the repulsive core of the NN potential.  

\section{Operator Product Expansion and potentials at short distance}

\subsection{General argument}
\label{sec:general}
Let us generalize the argument in the introduction, which relates 
the operator product expansion(OPE) to the short distance behavior 
of the potential.
We write the equal time Bethe-Salpeter (BS) wave function as
\begin{eqnarray}
\varphi^{E}_{AB}(\vec r) &=& \langle 0 \vert O_A(\vec r, 0) O_B(\vec 0, 0) \vert E  \rangle
\end{eqnarray}
where $\vert E \rangle$ is the eigen-state of the system with the energy $E$, 
and $O_A, O_B$ are some operators of the system. Here we suppress further 
quantum numbers of the state $\vert E\rangle $ other than $E$ for simplicity. 
The OPE of $O_A$ and $O_B$ is written as
\begin{eqnarray}
O_A(\vec r,0) O_B(\vec 0,0)&\simeq& \sum_C D_{AB}^C(\vec r) O_C(\vec 0, 0),
\quad r=\vert \vec r\vert \rightarrow 0.
\end{eqnarray}
We here assume that the coefficient function $D_{AB}^C$  behaves as
\begin{eqnarray}
D_{AB}^C(\vec r) &\simeq& ( - \log r)^{\beta_{AB}^C}
\end{eqnarray}
if $O_C$ has the same mass dimension of $O_A O_B$.
Therefore the BS wave function becomes
\begin{eqnarray}
\varphi_{AB}^E(\vec x) \simeq \sum_C (-\log r)^{\beta_{AB}^C} D_C(E),
\end{eqnarray}
where $ D_C(E) = \langle 0 \vert O_C(\vec 0,0)\vert E\rangle$.

In the approach of Ref.\cite{IAH, AHI1, AHI2}, the potential is defined 
by the BS wave function through the Schr\"odinger equation as
\begin{eqnarray}
V(\vec r) = E +\frac{1}{2\mu}\frac{\nabla^2 \varphi_{AB}^E(\vec r)}{\varphi_{AB}^E(\vec r)},
\end{eqnarray}
where $\mu$ is the reduced mass. Since
\begin{eqnarray}
\nabla^2 &=& \frac{1}{r^2} \frac{d}{d r}r^2\frac{d}{dr} - \frac{L(L+1)}{r^2}
\end{eqnarray}
for the state with angular momentum $L$, the dominant contribution of 
the potential for non-zero $L$ at short distance is trivially given by
\begin{eqnarray}
V_L( r) \simeq -\frac{L(L+1)}{r^2} +\cdots .
\end{eqnarray}
As a non-trivial case let us consider the $L=0$ case. 
There are two cases, depending on the maximum value of $\beta_{AB}^C$ 
defined as $\beta_X = \displaystyle\max_{C}\, \beta_{AB}^C$. 
\begin{enumerate}
\item[(1)] $\beta_X \not = 0$:  In this case, the potential at short distance is universally given by
\begin{eqnarray}
V(r) \simeq -\frac{\beta_X}{r^2(-\log r)} .
\end{eqnarray}
If $\beta_X > 0$ the interaction is attractive at short distance, 
while it is repulsive if $\beta_X < 0$.
\item[(2)]  $\beta_X =0$: We denote $\beta_Y < 0$ is the second largest among $\beta_{AB}^C$'s. 
The potential at short distance becomes
\begin{eqnarray}
V(r) \simeq \frac{D_Y(E)}{D_X(E)} \frac{-\beta_Y}{r^2 (-\log r)^{1-\beta_Y}},
\end{eqnarray}
which is attractive for 
$D_Y(E)/D_X(E) < 0 $ while repulsive for $D_Y(E)/D_X(E) > 0 $.
\end{enumerate}

\subsection{OPE in QCD}
Since QCD is an asymptotically free theory, the 1-loop calculation for 
anomalous dimensions becomes exact at short distance. 
The OPE in QCD is written as
\begin{eqnarray}
O_A(\vec r, 0) O_B(\vec 0,0)&=&\sum_C D_{AB}^C(r,g,m,\mu)O_C(\vec 0,0)
\end{eqnarray}
where $g$ ($m$) is the renormalized coupling constant (quark mass) at scale $\mu$.
In the limit that $r  = \mathrm{e}^{-t} R \rightarrow 0 $ 
( $t\rightarrow\infty$ with fixed $R$),
the renormalization group analysis leads to
\begin{eqnarray}
D_{AB}^C(r,g,m,\mu)\mathop{\sim}_{r\to0} 
(-2\beta^{(1)} g^2\log r)^{\gamma_{AB}^{C,(1)}/(2\beta^{(1)})}
D_{AB}^C(R, 0,0,\mu),
\end{eqnarray}
where 
\begin{eqnarray}
\beta^{(1)} = \frac{1}{16\pi^2}\left(11-\frac{2N_f}{3}\right)
\end{eqnarray}
is the QCD 1-loop beta-function coefficient, and 
\begin{eqnarray}
\gamma_{AB}^{C,(1)} &=& \gamma_A^{(1)} + \gamma_B^{(1)} - \gamma_C^{(1)}.
\end{eqnarray}
Here $\gamma_X^{(1)}$ is the 1-loop anomalous dimension of the operator $O_X$. 
An appearance of $D_{AB}^C(R,0,0,\mu)$ on the right-hand side tells us that
it is enough to know the OPE only at tree level. 
From the above expression, $\beta^C_{AB}$ in the previous subsection 
is given by
\begin{eqnarray}
\beta^C_{AB} &=& \frac{\gamma_{AB}^{C,(1)}}{2\beta^{(1)}}.
\end{eqnarray}
Therefore our task is to calculate $\gamma_X^{(1)}$ for 
3 and 6--quark operators.

\section{Anomalous dimensions for 6--quark operators}
\subsection{6 quark operators}
The OPE of two baryon operators at tree level is given by
\begin{eqnarray}
B_1(x) B_2(0) &=& B_1(0) B_2(0) + x^\mu (\partial_\mu B_1(0)) B_2(0) +
\frac12 x^\mu x^\nu (\partial_\mu\partial_\nu B_1(0) ) B_2(0)+ \cdots
\end{eqnarray}
where the first term corresponds to the $L=0$ contribution, 
the second to the $L=1$, and so on.
In this report we consider the $L=0$ case (the first term) only.
We denote the general form of a gauge invariant 3--quark operator as 
\begin{eqnarray}
B_{\alpha\beta\gamma}^{fgh}(x)\equiv B_\Gamma^F(x) =\varepsilon^{abc} q_\alpha^{a,f}(x)
q_\beta^{b,g}(x) q_\gamma^{c,h}(x)
\end{eqnarray}
where $\alpha,\beta,\gamma$ are spinor, $f,g,h$ are flavor, 
$a,b,c$ are color indices of quark field $q$. The 6--quark operator 
is constructed from two 3--quark operators as
\begin{eqnarray}
B_{\Gamma_1,\Gamma_2}^{F_1,F_2}(x) = B_{\Gamma_1}^{F_1}(x) B_{\Gamma_2}^{F_2}(x)
\end{eqnarray}
where $\Gamma_i = \alpha_i\beta_i\gamma_i$ and $F_i =f_ig_ih_i$ ($i=1,2$).
Since quarks are fermions, there are some linear dependencies among 6--quark 
operators. We have to determine a set of independent 6--quark operators. 
It is not so easy, however, to find them using a quark field basis. 
Instead we use a simpler method mentioned below.

As the choice of the gauge fixing in perturbation theory, 
we take the covariant gauge with gauge parameter $\lambda$. 
Since both the 3--quark operator $B_\Gamma^F$ and 6--quark operator 
$B_{\Gamma_1,\Gamma_2}^{F_1,F_2}$ are gauge invariant, 
$\gamma_{AB}^{C,(1)}$ must be independent of the gauge parameter $\lambda$. 
The $\lambda$ dependent term in the calculation of $\gamma_{AB}^{C,(1)}$ 
at 1-loop becomes
\begin{eqnarray}
\lambda \left( 9 B_{\Gamma_1,\Gamma_2}^{F_1,F_2} 
+ 3 \sum_{i,j=1}^3 B_{( \Gamma_1,\Gamma_2)[ij]}^{(F_1,F_2)[ij] } \right),
\label{eq:constraint}
\end{eqnarray}
where the $i$-th index of $abc$ and the $j$-th index of $def$ is interchanged 
in $(abc,def)[ij]$. For example, 
$(\Gamma_1,\Gamma_2)[11] = \alpha_2\beta_1\gamma_1,\alpha_1\beta_2\gamma_2$ 
or $(\Gamma_1,\Gamma_2)[21]
= \alpha_1\alpha_2\gamma_1,\beta_1\beta_2\gamma_2$. Note that the 
interchange occurs simultaneously for both 
$\Gamma_1,\Gamma_2$ and $F_1,F_2$ in the above formula.
The gauge invariance implies eq.(\ref{eq:constraint}) $= 0$, 
which yields constraints among the 6--quark operators. 
For example, let us consider the case that 
$\Gamma_1,\Gamma_2 =\alpha\alpha\beta,\alpha\beta\beta$ 
and $F_1,F_2 =ffg,ffg$, for which the constraint becomes
\begin{eqnarray}
3\left( 3 B_{\alpha\alpha\beta,\alpha\beta\beta}^{ffg,ffg} + (3-2)B_{\alpha\alpha\beta,\alpha\beta\beta}^{ffg,ffg} +B_{\alpha\alpha\alpha,\beta\beta\beta}^{fff,fgg}
+(2-1) B_{\alpha\beta\beta,\alpha\alpha\beta}^{fgg,fff} \right)&=& 0 \nonumber \\
\Rightarrow  4 B_{\alpha\alpha\beta,\alpha\beta\beta}^{ffg,ffg}+ 
B_{\alpha\alpha\alpha,\beta\beta\beta}^{fff,fgg}+ 
B_{\alpha\beta\beta,\alpha\alpha\beta}^{fgg,fff}  &=& 0,
\end{eqnarray}
where minus signs in the first line come from the property that 
$B_{\Gamma_2,\Gamma_1}^{F_2,F_1}= - B _{\Gamma_1,\Gamma_2}^{F_1,F_2}$. 
There are no further relations among 6--quark operators beyond 
(\ref{eq:constraint}).

\subsection{1-loop contributions}
\label{sec:1-loop}
Only the divergent part of the gauge invariant contribution at 1-loop 
is necessary to calculate the anomalous dimension of 6--quark operators 
at 1-loop.
The building block of 1-loop calculations is the gluon exchange between 
two quark lines. 
If both two quark lines belong to one operator, 
either $B_{\Gamma_1}^{F_1}$ or  $B_{\Gamma_2}^{F_2}$, 
the contribution is canceled by the renormalization factor of the 3--quark 
operator $B_{\Gamma_i}^{F_i}$, so that the divergent term does not 
contribute to 
$\gamma_{AB}^{C,(1)} =\gamma_A^{(1)} + \gamma_B^{(1)} - \gamma_C^{(1)}$.
If one quark line comes from $B_{\Gamma_1}^{F_1}$ and the other 
from $B_{\Gamma_2}^{F_2}$, the divergent term contributes to 
$\gamma_{AB}^{C,(1)}$.
Suppose that one quark line has indices $(\alpha_A, f_A)$ at one end 
and $(\alpha_1,f_1)$ at the other end and the other quark line has 
$(\alpha_B, f_B)$  and $(\alpha_2,f_2)$. The divergent contribution 
from the 1-gluon exchange can be expressed as
\begin{eqnarray}
\frac{g^2}{96\pi^2} \frac{1}{\bar\epsilon} 
\left\{
\delta_{f_1f_A}\delta_{f_2f_B}\left[\delta_{\alpha_1\alpha_A}\delta_{\alpha_2\alpha_B}
-2\delta_{\alpha_2\alpha_A}\delta_{\alpha_1\alpha_B}\right]
+ 3 \delta_{f_2f_A}\delta_{f_1f_B} \left[\delta_{\alpha_2\alpha_A}\delta_{\alpha_1\alpha_B}
-2\delta_{\alpha_1\alpha_A}\delta_{\alpha_2\alpha_B}\right] \right\}
\end{eqnarray}
for $(\alpha_1,\alpha_2)\in (R,R)$ or $(\alpha_1,\alpha_2)\in (L,L)$, 
where $R$ and $L$ means the right and the left handed component of the 
spinor indices. Other combinations, $(R,L)$ or $(L,R)$, vanish. 
This property comes from the fact that the gluon coupling to quarks 
is chirally symmetric and the chirally non-symmetric quark mass term does not 
contribute to the divergence.

\subsection{Chiral decomposition of 6--quark operators}
The physical nucleon operators are constructed from general 3--quark 
operators as
\begin{eqnarray}
B_\alpha^f &=& B_{\alpha\beta\gamma}^{fgh} \left ( C\gamma_5\right)_{\beta\gamma} (i\tau_2)^{gh}
\end{eqnarray}
where $C$ is the charge conjugation matrix, $f,g,h $ are $u$ or $d$,  
and $\tau_2$ is the Pauli matrix in the flavor space.  
The spinor index $\alpha$ is restricted to the positive energy component 
such that $\alpha=1,2$ in the Dirac representation of the $\gamma$ matrices.

In this report we consider $L=0$ two nucleon states, which are $^1S_0$ 
and $^3S_1$.
Here we use the notation $^{2S+1} L_J$ where $S$ is the total spin, 
$L$ is the orbital angular momentum and $J$ is the total angular momentum.
The 6--quark operator for $^1S_0$, which is the spin-singlet and 
isospin-triplet state, and for $^3S_1$ (the spin-triplet and isospin-singlet 
state) are given by
\begin{eqnarray}
BB(^1S_0) = (i\sigma_2)_{\alpha\beta} B_{\alpha}^f B_{\beta}^f, \qquad
BB(^3S_1) = (i\tau_2)^{fg} B_{\alpha}^f B_{\alpha}^g,
\end{eqnarray}
where the summation is taken for the repeated index.
Both 6--quark operators have the following chiral decomposition:
\begin{eqnarray}
BB &=& B_{LL} B_{LL} +B_{LL} B_{LR}+ B_{LL} B_{RL}+B_{LR} B_{LR}
+B_{LL} B_{RR} +\left( L\leftrightarrow R\right)
\end{eqnarray}
where $B_{XY}$ means $B_{\alpha,[\beta\gamma]}$ with $\alpha\in X$ and
$[\beta,\gamma]\in Y$ for $X,Y = R$ or $L$.

\subsection{Anomalous dimensions}
We now give our main results in this report. We define
\begin{eqnarray}
\gamma_{AB}^{C,(1)} &=&\gamma_A^{(1)}+\gamma_B^{(1)}- \gamma_C^{(1)} \equiv
\frac{1}{32\pi^2} \gamma.
\end{eqnarray}
The eigen-operators of the anomalous dimension matrix $\gamma$ are found 
to correspond to the chirally decomposed operators in the previous subsection.
We give the eigenvalue of each operator in table \ref{tab:gamma}, 
which shows that the operator with zero anomalous dimension always 
exists and other anomalous dimensions are all negative for both 
$^1S_0$ and $^3S_1$ states. This corresponds to the case (2) of 
the general discussion in the section \ref{sec:general}:
\begin{itemize}
\item[(2)] ($\beta_X =0$, $\beta_Y = -3/(33-2N_f)$) for $^1S_0$ and ($\beta_X=0$,  $\beta_Y = -1/(33-2N_f)$) for $^3S_1$. 
\end{itemize}

The appearance of zero eigenvalues in both $^1S_0$ and $^3S_1$ states can 
be understood as follows. As mentioned in section \ref{sec:1-loop}, 
1-loop contributions to the $\gamma_{AB}^{C,(1)}$ exist only if spinor 
indices of two quark lines, one from $\Gamma_1$ the other from $\Gamma_2$ 
in $B_{\Gamma_1} B_{\Gamma_2}$, belong to the same chirality (left or right). 
Since $B_{LL} B_{RR} + B_{RR}B_{LL}$ has no such combination, 
$ \gamma_{AB}^{C,(1)}$ is always zero for this type of operators. 
As pointed out before, this property is the consequence of the 
chiral symmetry in QCD interactions.

\begin{table}[tbh]
\caption{The eigenvalue $\gamma$ for anomalous dimension of each eigen operator in $^1S_0$ and $^3S_1$ states. In the table $(XY,ZW)$ means $B_{XY}B_{ZW}+(R\leftrightarrow L)$.}
\begin{center}
\begin{tabular}{|c|cccccc|}
\hline
 & $(LL,LL)$ & $(LL,LR)$ & $(LL,RL)$ &$(LR,LR)$ &
$(LR,RL)$ & $(LL,RR)$ \\
\hline
$\gamma(^1S_0)$ & $-12$ & $-4$ & $-8$ & $-8$ & $-6$ & $0$ \\
$\gamma(^3S_1)$ & $-28/3$ & $-4/3$ & $-8$ & $-16/3$ & $-6$ & $0$ \\
\hline
\end{tabular}
\label{tab:gamma}
\end{center}
\end{table}

\section{Conclusion}
The OPE and renormalization group analysis in QCD predicts the 
universal functional form of the nucleon-nucleon potential at 
short distance:
\begin{eqnarray}
V(r) &\simeq& \frac{D_Y(E)}{D_X(E)} \frac{-\beta_Y}{r^2(-\log r)^{1-\beta_Y}},
\qquad r\rightarrow 0,
\end{eqnarray}
which is a little weaker than a $1/r^2$ singularity. We obtain
\begin{eqnarray}
\beta_Y(^1S_0) =-\frac{3}{33-2N_f}, \qquad
\beta_Y(^3S_1) =-\frac{1}{33-2N_f}\,.
\end{eqnarray}

The anomalous dimension spectrum, however, cannot alone tell whether 
the potential at short distance is repulsive or attractive.
If we evaluate $D_X(E)$ and $D_Y(E)$ by the chiral effective theory 
at the leading order ({\it i.e.} the tree level), we obtain
\begin{eqnarray}
\frac{D_Y(E)}{D_X(E)}(^1S_0) &=& \frac{2 E}{m_N}, \qquad
\frac{D_Y(E)}{D_X(E)}(^3S_1) = \frac{2m_N}{E},
\end{eqnarray}
where $E = \sqrt{\vec p^2 + m_N^2}$ and $\vec p$ is the relative 
momentum of two nucleons.
We find a repulsive core for both states. In particular, in the low energy 
region such that $\vec p^2 \ll m_N^2$, the repulsive potential at 
short distance is almost energy independent, since
\begin{eqnarray}
\frac{D_Y(E)}{D_X(E)}(^1S_0)  &\simeq& \frac{D_Y(E)}{D_X(E)}(^3S_1) \simeq 2.
\end{eqnarray}
Furthermore no low energy constant of the effective theory is required 
at leading order to obtain the above result.

There are several interesting extensions of the analysis using the OPE. 
The extension of the OPE analysis to the 3--flavor case may reveal the 
nature of the repulsive core in the baryon-baryon potentials. 
Since quark mass can be neglected in this OPE analysis, the calculation 
can be done in the exact SU(3) symmetric limit. It is also interesting 
to investigate the existence or the absence of the repulsive core 
in the 3--body nucleon potential. In this case, we have to calculate 
anomalous dimensions of 9--quark operators at 2-loop level.
More precise evaluations of the matrix element 
$D_X(E) = \langle 0 \vert O_X\vert E\rangle$ would be preferable. 
The most straightforward extension is to analyze the tensor force 
and $LS$ force by the OPE. Preliminary results indicate that
\begin{eqnarray}
V_T(r) &\simeq& C_0 + C_1 (-\log r)^{\beta_Y-1} , \qquad
V_{LS}(r) \simeq  -\frac{12}{m_N r^2},
\end{eqnarray}
where $LS$ force has the strong attractive core at short distance, 
while no repulsive core exists for the tensor potential. 
Absence of the repulsive core predicted in the tensor force is consistent 
with recent numerical simulations\cite{AHI2, IAH2}. 

\vspace{-0.2cm}
\section*{Acknowledgments}
S. A. is supported in part by Grant-in-Aid of the Ministry of Education, 
Sciences and Technology, 
Sports and Culture (Nos. 20340047, 20105001, 20105003).
This investigation was supported in part by the Hungarian National 
Science Fund OTKA (under T77400).

\end{document}